# Clustering, Encoding and Diameter Computation Algorithms for Multidimensional Data

Mugurel Ionuţ Andreica, Eliana-Dina Tîrşa
Politehnica University of Bucharest, Computer Science Department, {mugurel.andreica, eliana.tirsa}@cs.pub.ro

*Abstract*-In this paper we present novel algorithms for several multidimensional data processing problems. We consider problems related to the computation of restricted clusters and of the diameter of a set of points using a new distance function. We also consider two string (1D data) processing problems, regarding an optimal encoding method and the computation of the number of occurrences of a substring within a string generated by a grammar. The algorithms have been thoroughly analyzed from a theoretical point of view and some of them have also been evaluated experimentally.

## I. INTRODUCTION

In this paper we consider several multidimensional data processing problems, for which we present novel algorithmic techniques. Two important problems in processing multidimensional sets of points are: clustering and the computation of the diameter with respect to a distance function. An efficient clustering method has two advantages: it provides information about the structure of the set of points and can be used for providing a more compact representation of the set of points (by describing just the clusters and not the individual points). Computing the diameter of a set of points is important particularly in association with other metrics. It can provide relevant information regarding the distribution of the points in space.

One-dimensional data (e.g. textual data) is a particular case of multidimensional data which is of a high practical interest. Computing optimal encodings and analyzing large amounts of data (given explicitly or implicitly) are important in many practical fields. In this paper we propose a new algorithm for computing an optimal encoding with respect to several rules, and a solution for computing the number of occurrences of a substring in a large string (implicitly generated by a grammar).

The rest of this paper is structured as follows. In Section II we present related work. In Section III we discuss the problem of covering a set of points by using at most *kh* (restricted) hyper-rectangles. A hyper-rectangle is the simplest way of representing a multi-dimensional cluster. The fact that the hyper-rectangles need not be non-overlapping implies that the clusters need not be disjoint. In Section IV we present a new algorithm for computing the diameter of a set of points, using a new distance function. In Section V we discuss two string processing problems, regarding the optimal encoding and the analysis of text data. In Section VI we conclude and discuss future work.

## II. RELATED WORK

Geometric K-center problems (related to the hyper-rectangle covering problem from Section III) were presented in [1]. Efficient algorithms for geometric optimization problems similar to those introduced in this paper were given in [2]. Multidimensional data structures like the range tree or multidimensional versions of 2D data structures like the segment tree were presented in [3, 4, 5]. A similar model for partitioning the dimensions into groups as the one we use in Section III, but applied to a multidimensional range minimum query problem, was mentioned in [5]. Related 2D data structures and related problems in multiple dimensions in the OLAP domain were considered in [6, 7].

## III. HYPER-RECTANGLE COVERING

### A. Problem Description and Algorithmic Solutions

We consider the following problem. We are given a set of $r$ points $(xc(i,1), ..., xc(i,d))$ ($1 \leq i \leq r$) in a d-dimensional space. We want to cover all of the points with at most $kh$ hyper-rectangles having the following properties. A hyper-rectangle is defined by $d$ intervals $[a_1,b_1], ..., [a_d, b_d]$ and contains all the points $(x_1, ..., x_d)$ for which $a_i \leq x_i \leq b_i$ ($1 \leq i \leq d$). The side length $len(i)$ of a hyper-rectangle in dimension $i$ is $b_i - a_i$. We consider the $d$ dimensions as being classified into $e \leq d$ groups. Let $g(i)$ be the group to which the dimension $i$ belongs ($1 \leq g(i) \leq e$; $1 \leq i \leq d$). The side lengths of the hyper-rectangles we want to place in each dimension $i$ must satisfy the constraint: $len(i)=f(i) \cdot l(g(i))$. Thus, the side lengths of each hyper-rectangle are uniquely defined by the values $l(j)$ ($1 \leq j \leq e$). Moreover, for each group $j$ we have a lower bound $lmin(j)$ and an upper bound $lmax(j)$, meaning that $lmin(j) \leq l(j) \leq lmax(j)$.

A hyper-rectangle can be placed anywhere in space and a point is covered by the (at most) $kh$ hyper-rectangles if it is contained within at least one of the hyper-rectangles. We want to place the (at most) $kh$ hyper-rectangles such that an aggregate function *aggf* of their costs is minimized (e.g. *aggf*=+). The cost of a hyper-rectangle can be any non-negative value which depends on the side lengths of the hyper-rectangle and, possibly, on the points contained within it, but must not depend on the actual coordinates of the hyper-rectangle. The cost function should be non-decreasing with respect to each side length. The aggregate value must be computed using a commutative function of the hyper-rectangles' costs, which must be non-decreasing with respect to the "addition" of the cost of a new hyper-rectangle.

We will present several solutions for this problem, each of them successively improving the previous one. We will start by sorting the coordinates of the points in each dimension. Let $xo(i,1) \leq xo(i,2) \leq ... \leq xo(i,r)$ be the order of the points in the dimension $i$. We will remove the duplicates in the ordering (maintaining only one coordinate with a given value) and we obtain the ordering: $xp(i,1) < xp(i,2) < ... < xp(i,m(i))$, where $m(i)$ is the number of distinct coordinates in the dimension $i$. We will denote by $n=max\{m(i)|1{\leq}i{\leq}d\}$. We will use the value of $n$ when analyzing the time complexity of our solutions, because it will be easier than using the values $m(i)$ explicitly. If the point set is sparse, then we may have $n=O(r)$. If the point set is dense, then we may have $n=O(r^{1/d})$.

The first step in each of the presented solutions is to compute the minimum bounding hyper-rectangle (MBR) of the $r$ points. This hyper-rectangle $[xp(i,1),xp(i,m(i))]$ $(1{\leq}i{\leq}d)$ has the minimum (hyper-)volume possible and still contains all the $r$ points inside. For simplicity, we will assume that the parameters of all the hyper-rectangles will be expressed in terms of indices into the sorted arrays $xp(*)$, i.e. $a_i=j$, $b_i=k$ means that the side of the hyper-rectangle is $[xp(i,j), xp(i,k)]$.

The first presented solution is a generic, yet naïve, solution and is described in the following pseudo-code:

**hrcover(S, HRMBR, kh):**
**if** $(|S|=0)$ **then return** $0$
**if** $(kh=1)$ **then** {
  $xmin(j)=xp(j,HRMBR.a_j)$ ; $xmax(j)=xp(j,HRMBR.b_j)$ $(1{\leq}j{\leq}d)$
  $l(j)=lmin(j)$ $(1{\leq}j{\leq}e)$
  **for** $i=1$ **to** $d$ **do** $l(g(i))=max(l(g(i)), (xmax(i)-xmin(i))/f(i))$
  **if** $(l(j)>lmax(j))$ (for some $1{\leq}j{\leq}e$) **then return** $+\infty$
  $xmax(j)=xmin(j)+f(j){\cdot}l(g(j))$ $(1{\leq}j{\leq}d)$
  **return cost**$([xmin(j),xmax(j)]$ $(1{\leq}j{\leq}d), S)$
} **else** {
  $C_{min}=+\infty$
  **for each** $HR$ **in** *generateHyperRectangles(HRMBR, kh)* **do** {
    $U=\{p$ in $S$ | $HR.xmin(j){\leq}xc(p,j){\leq}HR.xmax(j)$ for every $1{\leq}j{\leq}d\}$
    $S'=S{\setminus}U$
    $C=$**cost**$([HR.xmin(j), HR.xmax(j)]$ $(1{\leq}j{\leq}d), U)$
    $MBR'=$**computeMBR**$(S')$
    $C_{min}=min\{C_{min}, aggf(C,$ **hrcover**$(S', MBR', kh-1))\}$ }
  **return** $C_{min}$ }
**generateHyperRectangles(HRMBR, kh):**
  $HRList = \{\}$
  **for each tuple** $(c_1, ..., c_d)$ **such that** $HRMBR.a_i{\leq}c_i{\leq}HRMBR.b_i$ $(1{\leq}i{\leq}d)$ **do** {
    $xmin(j)=xp(j,c_j)$ $(1{\leq}j{\leq}d)$
    **let** $difSet(g_i)=\{(xp(j,a)-xp(j,c_j))/f(j)$ | $a{\geq}c_j$; $g(j)=g_i$; $1{\leq}j{\leq}d\}$ $(1{\leq}g_i{\leq}e)$
    **let** $difMax(g_j)=max\{dif$ | $dif$ in $difSet(g_j)\}$ $(1{\leq}g_j{\leq}e)$
    **for each tuple** $(l(1), ..., l(e))$ **such that** $lmin(j){\leq}l(j){\leq}lmax(j)$ **and** $((l(j)$ in $difSet(j))$ **or** $(l(j)=lmin(j)>difMax(j)))$ $(1{\leq}j{\leq}e)$ **do** {
      $xmax(j)=xmin(j)+f(j){\cdot}l(g(j))$ $(1{\leq}j{\leq}d)$
      $HRList.add([xmin(j),xmax(j)]$ $(1{\leq}j{\leq}d))$ }}
  **return** $HRList$

The parameters of the **hrcover** algorithm are $S=$the set of yet uncovered points, $HRMBR=$the minimum bounding hyper-rectangle of the points in $S$ (expressed in terms of indices in the sorted arrays $xp(*)$) and $kh=$the number of remaining hyper-rectangles. The **cost** function takes as arguments the current hyper-rectangle and the set of points $U$ located inside it. If the cost function does not depend on the set $U$, then it can be computed in $O(1)$ time; otherwise, its time complexity may be higher (depending on the actual function). We will denote by $CC(r)$ the time complexity of computing the cost function when $|U|=O(r)$. **generateHyperRectangles(HRMBR, kh)** generates a list of possible placements for the current hyper-rectangle. **computeMBR(S')** computes the minimum bounding hyper-rectangle $MBR'$ of the points from the set $S'$. The parameters of $MBR'$ will also be expressed in terms of indices into the arrays $xp(*)$. In order to compute these indices, we will store for each point $i$ and dimension $j$ the index $idx(i,j)=k$ such that $xp(j,k)=xc(i,j)$ (this index can be computed during the initial sorting of the coordinates along each dimension). Thus, $MBR'$ can be computed in $O(|S'|{\cdot}d)$ time (we will maintain the point with the minimum and maximum coordinate in each dimension and then compute the result using the values $idx(*,*)$).

For each of the $kh$ hyper-rectangles, the algorithm considers each possible value of their left and right coordinates in each dimension. For each assignment of the coordinates, the algorithm computes the set of points outside of the hyper-rectangle and their minimum bounding hyper-rectangle. This takes $O(r{\cdot}d{\cdot}n^{2{\cdot}e})$ time for each of the first $kh-1$ hyper-rectangles. The final time complexity is $O(r^{kh-1}{\cdot}d^{kh-1}{\cdot}n^{2{\cdot}e{\cdot}(kh-1)}+r{\cdot}d{\cdot}log(r))$. This analysis is correct only if the cost function can be computed in $O(1)$ time. Otherwise, we need to include the time complexity of computing the cost function within the time complexity equation. For the general case, we will denote by $T(r,n,kh)=$the time complexity for computing an optimal cover of $O(r)$ points with at most $n$ distinct coordinates in each dimension and using at most $kh$ hyper-rectangles. $T(r,n,1)=O(d+CC(r))$. $T(r,n,kh{\geq}2)=O((r{\cdot}d+CC(r)){\cdot}n^{2{\cdot}e}{\cdot}T(r,n,kh-1))$. We must also add the initial $O(r{\cdot}d{\cdot}log(r))$ factor for sorting the coordinates and computing the initial minimum bounding hyper-rectangle.

The presented algorithm can be optimized in two directions. First, we can generate fewer potential hyper-rectangles. Second, we can compute in a more efficient manner the minimum bounding hyper-rectangle of the points not contained in any of the hyper-rectangles that were placed so far.

The first optimization is based on the following observation. The MBR of a set of points has $2{\cdot}d$ sides (it is defined by $2{\cdot}d$ parameters). If we want to cover the points by using $kh$ hyper-rectangles, then some of the parameters of these hyper-rectangles will need to coincide with the parameters of the MBR (or exceed them in the corresponding direction). Using Dirichlet's principle, there must be at least one hyper-rectangle with $q=2{\cdot}d/kh$ (rounded up) common parameters with the MBR. This hyper-rectangle will be placed next. We will consider all the combinations of $q$ common parameters and choose the remaining $2{\cdot}d-q$ parameters as before. The pseudocode of the optimized *generateHyperRectangles* function is below:

**generateHyperRectangles(HRMBR, kh):**
  $HRList = \{\}$
  $q = \left\lceil \dfrac{2 \cdot d}{kh} \right\rceil$
  **for each subset** $SQ$ **with** $q$ **elements of the set** $\{-1, -2, ..., -d, +1, +2, ..., +d\}$ **do**
    **for each tuple** $(ha_1, ..., ha_d, hb_1, ..., hb_d)$ **such that** $HRMBR.a_i{\leq}ha_i{\leq}hb_i{\leq}HRMBR.b_i$ $(1{\leq}i{\leq}d)$ **and** (**if** $-i$ **is in** $SQ$ **then** $ha_i=HRMBR.a_i$ **else** *true* $(1{\leq}i{\leq}d)$) **and** (**if** $+i$ **is in** $SQ$ **then** $hb_i=HRMBR.b_i$ **else** *true* $(1{\leq}i{\leq}d)$) **do** {
      $xmin(j)=xp(j,ha_j)$ $(1{\leq}j{\leq}d)$ ; $xmax(j)=xp(j,hb_j)$ $(1{\leq}j{\leq}d)$
      $l(j)=lmin(j)$ $(1{\leq}j{\leq}e)$

    **for** $i=1$ **to** $d$ **do** $l(g(i))=max\{l(g(i)), (xmax(i)-xmin(i))/f(i)\}$
    **if** $(l(j)>lmax(j))$ (for some $1\leq j\leq e$) **then continue**
    $xmax(j)=xmin(j)+f(j)\cdot l(g(j))$ $(1\leq j\leq d)$
    $HRList.add([xmin(j),xmax(j)]$ $(1\leq j\leq d))$ }
 **return** $HRList$

The time complexity of the algorithm using the optimized *generateHyperRectangles* function can be analyzed as follows. At each function call with $kh\geq 2$, $C(2\cdot d,q)$ sets $SQ$ are considered ($C(a,b)$=combinations of $a$ elements taken $b$ at a time). For each set $SQ$, $O(min\{n^{2\cdot e}, n^{2\cdot d\cdot q}\})$ hyper-rectangles are considered. Thus, the total number of hyper-rectangles considered at each step may decrease significantly.

The second improvement consists of the way the MBR $[a_i',b_i']$ $(1\leq i\leq d)$ of the points from the set $S'$ is computed. So far, we computed the MBR in $O(|S|)$ time. However, we can do even better. We will insert all the points from the set $S$ into several data structures $DS_{j,k}$. Then, after choosing the parameters $[xmin(j),xmax(j)]$ $(1\leq i\leq d)$ of the current hyper-rectangle, we will proceed as follows. We need to compute $a_i'=min(W_i=\{idx(j, i) \mid j$ in $S$ and $(xc(j,k)<xmin(k)$ (for some value of $k$; $1\leq k\leq d$) or $xc(j,k)>xmax(k)$ (for some value of $k$; $1\leq k\leq d$) )}) and $b_i'=max(W_i)$ $(1\leq i\leq d)$. Each possibility $xc(j,k)<xmin(k)$ $(xc(j,k)>xmax(k))$ defines an orthogonal half-space. We need to query $DS_{*,*}$ with respect to such a half-space. This is a special case of orthogonal range query.

We will construct $d^2$ data structures $DS_{j,k}$ $(1\leq j\leq d)$ into which every point $i$ from $S$ is inserted, with a weight equal to $idx(i,k)$. Then, we need to answer orthogonal range minimum and maximum queries over these data structures. An orthogonal range min (max) query returns the minimum (maximum) weight of a point located in the query range. Then, $a_i'=min_{1\leq k\leq 2\cdot d}\{W(i,k)=\{idx(q,i) \mid j=(k+1)/2$ (if $k$ is odd) or $k/2$ (if $k$ is even), $xc(q,j)<xmin(j)$ (if $k$ is odd) or $xc(q,j)>xmax(j)$ (if $k$ is even)}} and $b'(i)=max_{1\leq k\leq 2\cdot d}\{W(i,k)\}$.

Let $r'=|S|$. The $DS_{j,k}$ data structure simply sorts the points according to their $xc(*,j)$ coordinates, obtaining an ordering $xc(p_{j,k}(1),j)\leq\ldots\leq xc(p_{j,k}(r'),j)$. We will also have $xc(p_{j,k}(0)=0,j)=-\infty$ and $xc(p_{j,k}(r'+1)=r+1,j)=+\infty$. Then, we compute a prefix minimum $pmin_{j,k}$ and a prefix maximum $pmax_{j,k}$. $pmin_{j,k}(0)=+\infty$ and $pmax_{j,k}(0)=-\infty$. For $1\leq i\leq r'$, $pmin_{j,k}(i)=min\{idx(p_{j,k}(i),k), pmin_{j,k}(i-1)\}$ and $pmax_{j,k}(i)=max\{idx(p_{j,k}(i),k), pmax_{j,k}(i-1)\}$. Similarly, we compute a suffix minimum and a suffix maximum, $smin_{j,k}$ and $smax_{j,k}$. $smin_{j,k}(r'+1)=+\infty$ and $smax_{j,k}(r'+1)=-\infty$. For $1\leq i\leq r'$, $smin_{j,k}(i)=min\{idx(p_{j,k}(i),k), smin_{j,k}(i+1)\}$ and $smax_{j,k}(i)=max\{idx(p_{j,k}(i),k), smax_{j,k}(i+1)\}$. The queries for the minimum (maximum) value $idx(q,k)$ from the half-space $xc(q,j)<xmin(j)$ $[xc(q,j)>xmax(j)]$ are answered by using the data structure $DS_{j,k}$. We binary search the maximum position $pos$ $(0\leq pos\leq r')$ such that $xc(p_{j,k}(pos),j)<xmin(j)$ [the minimum position $pos$ $(1\leq pos\leq r'+1)$ such that $xc(p_{j,k}(pos),j)>xmax(j)$] and return $pmin_{j,k}(pos)$ $(pmax_{j,k}(pos))$ [ $smin_{j,k}(pos)$ $(smax_{j,k}(pos))$ ]. If the values $xmin(j)$ $(xmax(j))$ which can be used as query parameters are known in advance and their number is not too large, then we can sort all the possible values and, for each such value, precompute the corresponding value $pos$: we traverse the set of sorted values in ascending (descending) order and, when searching for the corresponding value $pos$, we start from the value $pos$ found for the previously considered value and increase (decrease) it as long as the required condition holds; for the first considered value we start with $pos=0$ $(pos=r'+1)$. Then, we insert into a hash table $Ha$ $(Hb)$ each possible value $xmin(j)$ $(xmax(j))$ together with its corresponding $pos$ value and, when a query with parameter $xmin(j)$ $(xmax(j))$ is asked, we simply retrieve the value $pos$ from the hash table in $O(1)$ time.

Answering such queries takes $O(log(|S|))$ time per query (or $O(1)$ if we can use the hash tables). Thus, computing the MBR takes $O(d^2\cdot log(|S|))$ time (or only $O(d^2)$ time). Constructing the data structures takes $O(d^2\cdot r\cdot log(r))$ time overall. However, since the points can be sorted in the beginning according to each dimension, we can construct each data structure in $O(d^2\cdot r)$ time (by considering the corresponding sorted order and maintaining only those points which are still in $S$).

If the set of points is dense, then the number of distinct coordinates in each dimension is small. In this case, we can sort the points according to each dimension $j$ using a variation of count-sort. For each possible value $q$ of the coordinates in dimension $j$, we maintain a list $L_j(q)$ in which all the points $o$ with the coordinate $xc(o,j)=q$ are inserted (each insertion is performed in $O(1)$ time). Then, in order to obtain the final order of the points according to the dimension $j$, we simply concatenate the lists $L_j(q)$ in increasing order of the coordinate values $q$. This way, the sorting of the points in the dimension $j$ takes $O(n+r)$ time. We can use the same procedure for sorting the possible query parameters $xmin(j)$ $(xmax(j))$ of the data structures $DS_{j,*}$ when they are known in advance and their number is not too large.

The improvement regarding the use of data structures makes sense mostly when $kh=2$, because in this case we do not need to also compute the set $S'$ (since the obtained MBR is used directly at $kh=1$). If we also need to compute the set $S'$, then we will first compute the set $S'$ and only afterwards will we compute the MBR, in $O(|S'|)$ time, by considering every point from $S'$ (i.e. in the normal manner). However, if $|S'|$ is significantly smaller than $|S|$, we could proceed as follows. We could insert (initially) all the points from $S$ into a range tree $RT$. Then, the set $S'$ can be computed by reporting (rather than counting) all the points from $RT$ which are within a union of orthogonal query ranges. There are $3^d-1$ such ranges. The interval of each dimension $j$ is split into $3$ ranges: $[-\infty, HR.xmin(j))$, $[HR.xmin(j), HR.xmax(j)]$ and $(HR.xmax(j), +\infty]$, thus obtaining a division of the space into $3^d$ disjoint orthogonal regions. The only region we are not interested in is the "middle" region (i.e. the one corresponding to the current hyper-rectangle: $[HR.xmin(j), HR.xmax(j)]$ $(1\leq j\leq d)$). With these improvements, the time complexity of the algorithm decreases significantly.

Note that the idea of dividing the space into $3^d$ regions can also be used for computing the MBR. Instead of the $DS_{*,*}$ data structures, we will use $d$ d-dimensional range trees $RTDS_k$ $(1\leq k\leq d)$. In each range tree $RTDS_k$ we will insert all the points $i$ from $S$, and we associate to them a weight equal to $idx(i,k)$ $(1\leq k\leq d)$. Then, when computing the MBR, we need the following values: $a'_k=min\{RTDS_k.rangeMin(reg) \mid reg$ is one of the $3^d-1$ regions (i.e. not the middle region)} and $RTDS_k.rangeMin(reg)$ returns the minimum weight of a point located in the (orthogonal) region *reg* (*1≤k≤d*). Similarly, we have $b'_k$=*max{RTDS$_k$.rangeMax(reg) | reg is one of the $3^d$-1 regions (i.e. not the middle region)}* and *RTDS$_k$.rangeMax(reg)* returns the maximum weight of a point located in the (orthogonal) region *reg* (*1≤k≤d*). In this case, computing the MBR takes $O(d \cdot 3^d \cdot log^d(|S|))$ time, which is worse than the previous solution based on the $DS_{*,*}$ data structures.

**hrcoverOpt(S, HRMBR, kh):**
**if** *(|S|=0)* **then return** *0*
**if** *(kh=1)* **then {**
  *xmin(j)=xp(j,HRMBR.a$_j$); xmax(j)=xp(j,HRMBR.b$_j$) (1≤j≤d)*
  *l(j)=max{lmin(j), max{(xmax(i)-xmin(i))/f(i) | g(i)=j, 1≤i≤d}} (1≤j≤e)*
  **if** *(l(j)>lmax(j))* (for some *1≤j≤e*) **then return** *+∞*
  *xmax(j)=xmin(j)+f(j)·l(g(j)) (1≤j≤d)*
  **return cost(**[xmin(j),xmax(j)] (1≤j≤d), S**)**
**} else {**
  $C_{min}$=*+∞*
  *insert all the points from S in a d-dimensional range tree RT*
  *construct the $DS_{*,*}$ data structures*
  **for each** *HR* **in generateHyperRectangles(HRMBR, kh)* **do {**
    **if** *the cost function does not depend on the points located inside the hyper-rectangle* **then {**
      **if** *(kh>2)* **then** *construct S' as the union of $3^d$-1 range reporting queries in RT* (*)
      **else** *S'=any non-empty set of points (e.g. with just 1 point)*
      *C=**cost(**[HR.xmin(j), HR.xmax(j)] (1≤j≤d), -**)*
    **} else {**
      *U={p in S|HR.xmin(j)≤xc(p,j)≤HR.xmax(j) for every 1≤j≤d}*
      *S'=S\U*
      *C=**cost(**[HR.xmin(j), HR.xmax(j)] (1≤j≤d), U**) }*
    *MBR'=compute the MBR of the set S' using the data structures $DS_{*,*}$* (*)
    $C_{min}$=*min{$C_{min}$, aggf(C, **hrcoverOpt**(S', MBR', kh-1))}* **}**
  **return** $C_{min}$ **}**

The lines marked with (*) could be replaced by their "normal" counter-parts (i.e. the range tree *RT* could be ignored, in which case the set *S'* would be constructed as in the previous two solutions, or the MBR could be computed in *O(|S'|)* time). We may first count the number of points in *S'* (by summing the answers to the range count queries in *RT* for each of the $3^d$-1 regions) in order to decide how *S'* should be computed.

The two types of improvements we presented (fixing *q* of the parameters of the current hyper-rectangle to be identical to the MBR of the uncovered points and efficiently computing the MBR of the remaining uncovered points) are orthogonal to each other and can be used together or separately.

*B. Experimental Evaluation*

We randomly generated *80* dense point sets with *d=3*: *40* of them had *n=14*, and *40* of them had *n=40*. The number of points *r* varied from *0.1%* to *10%* of the value $n^3$. We considered *kh=2*, each dimension belonged to a separate group and we considered no bounds on the side lengths in each dimension. The cost function was the volume of a hyper-rectangle. We evaluated four algorithms: *A*-unoptimized hyper-rectangle generation and unoptimized MBR computation ; *B*-optimized hyper-rectangle generation and unoptimized MBR computation ; *C*-unoptimized hyper-rectangle generation and optimized MBR computation ; *D*-optimized hyper-rectangle generation and optimized MBR computation. The theoretical time complexities (for our test settings) of the four algorithms are: *A*-$O(n^9)$, *B*-$O(n^6)$, *C*-$O(n^6)$, *D*-$O(n^3)$. The algorithms were implemented in C++, compiled using the Visual Studio 2008 C++ compiler and were run on a 2 GHz processor running Windows Vista. The sums of running times (over the *40* test cases with the same value of *n*) are presented in Table I.

TABLE I
SUMS OF RUNNING TIMES FOR THE FOUR ALGORITHMS

|      | A          | B        | C        | D        |
|------|------------|----------|----------|----------|
| n=14 | 205 sec    | 8.12 sec | 2.22 sec | 0.13 sec |
| n=40 | > 86400 sec | 4684 sec | 954 sec  | 2.72 sec |

IV. DISTRIBUTION OF NODES IN A GEOMETRIC SPACE

In this section we make a proposal for a peer-to-peer message routing system, whose node identifiers are mapped into a metric space. Each node *q* of the system has an identifier which is a point *(x(q,1), …, x(q,d))* in a d-dimensional Euclidean space and is connected to at most *2·k·d* other peers (for each dimension *i*, *1≤i≤d*, the peer *q* is connected to the *k* peers *q'* with the smallest i-coordinates *x(q',i)* larger than or equal to *x(q,i)* and the *k* peers with the largest i-coordinates smaller than *x(q,i)*). In order to route a message from a source *s* to a destination *d*, the peers use the distance of the metric space of their identifiers and they forward the message to a neighboring peer which decreases the distance from the current peer to *d* (or minimizes it). Such a system poses many challenges which have to be analyzed, like scalability, fault tolerance, resistance to churn and appropriate distribution of the node identifiers in the metric space. We will consider the distribution problem in this section, using the following "distance" function: *dist((x(p,1),…,x(p,d)), (x(q,1), …, x(q,d)))=min{|x(p,i)-x(q,i)| 1≤i≤d}*. We want to compute the largest distance between any pair of points, which offers important information about the distribution of points in the metric space. Because the number of nodes in a peer-to-peer system can be quite large, we need an efficient solution (better than the trivial $O(n^2)$ solution which considers every pair of points).

We will binary search the maximum distance and we will perform a feasibility test on every candidate value $D_{cand}$ chosen by the binary search. If $D_{cand}$ is feasible, we will test a larger distance; otherwise, we will test a smaller one. For each point *i* *(x(i,1),…, x(i,d))*, we will compute the number of points *np(i)* located at distance at most $D_{cand}$ from it (including point *i* itself). If *np(i)<n* for some *i*, $D_{cand}$ is a feasible distance. We will preprocess the points into a d-dimensional range tree which can answer orthogonal range counting queries in $O(log^{d-1}(n))$ time (using fractional cascading). The number of points located at a distance $d≤D_{cand}$ from a point *i* is obtained by using the inclusion-exclusion principle on the answers to $2^d$-1 range count queries. A query *rcount(p,S)* is defined by a set *S* of dimensions for which the range is not unbounded (i.e. [-∞, +∞]); for each dimension *i* in *S*, the range is *[x(p,i)-$D_{cand}$, x(p,i)+$D_{cand}$]* and for the other dimensions not in *S* the range is unbounded. There are $2^d$-1 sets *S* (we exclude the empty set). *np(i)* is equal to the sum of $(-1)^{|S|-1}$·*rcount(i,S)* (*rcount(i,S)* is the answer to the range query defined by the set *S* and the point *i*). For $d=2$, $np(i)=count([-\infty, x(i,2)-D_{cand}], [+\infty, x(i,2)+D_{cand}]) + count([x(i,1)-D_{cand},-\infty], [x(i,1)+D_{cand},+\infty]) - count([x(i,1)-D_{cand}, x(i,2)-D_{cand}], [x(i,1)+D_{cand}, x(i,2)+D_{cand}])$. $count([xa,ya],[xb,yb])$ returns the number of points $(x_j,y_j)$ with $xa \leq x_j \leq xb$ and $ya \leq y_j \leq yb$. Thus, we can perform the feasibility test in $O(n \cdot log^{d-1}(n))$ time. The overall memory consumption of the range tree is $O(n \cdot log^{d-1}(n))$.

We can also use the following algorithm, which we will denote by *Sol(d)*, where *d* is the number of dimensions of the space. If $d=1$ then we only need to sort all the points according to their first (and only) coordinate and obtain an ordering $x(p(1),1) \leq \ldots \leq x(p(n),1)$. Then, we can compute the value $rcount(i,\{1\})$ either by binary searching the values $x(i,1)-D_{cand}$ and $x(i,1)+D_{cand}$ around the position $pos(i)$ of the point *i* in the sorted order (i.e. $p(pos(i))=i$), or by using a sweeping technique which will be described next for the more general case.

For $d \geq 2$ we will proceed as follows. Let's notice first that the values $rcount(*, S=\{j\})$ can be computed by sorting all the points according to their coordinates in the dimension *j* and then using the solution proposed for the case $d=1$, but considering that the only existing dimension is the dimension *j* ($1 \leq j \leq d$).

We will sort the points according to their $d^{th}$ coordinate and sweep them with a (d-1)-dimensional slab of infinite size in each dimension $1 \leq j \leq d-1$ and width $2 \cdot D_{cand}$ in the $d^{th}$ dimension. The position of the slab is denoted by its rightmost coordinate *xr* in dimension *d*. We will have three types of events for each point *i*:

1. the point *i* enters the slab; this event occurs when $xr=x(i,d)$
2. the point *i* is at the middle of the slab; this event occurs when $xr=x(i,d)+D_{cand}$
3. the point *i* leaves the slab; this event occurs when $xr=x(i,d)+2 \cdot D_{cand}$

The $3 \cdot n$ events will be sorted in ascending order of the value of *xr* when the event occurs and are then processed in this order. If multiple events occur at the same value of *xr*, then *entering* events take precedence over *middle* events, which take precedence over *leaving* events. When a point *i* enters the slab, we add $(x(i,1),\ldots,x(i,d-1))$ to a (d-1)-dimensional range tree *RT* constructed on the coordinates of the points in dimensions $1,\ldots,d-1$ (which is initially empty). When a point *i* leaves the slab, we remove $(x(i,1),\ldots,x(i,d-1))$ from the range tree. When a point *i* is exactly at the middle of the slab, we will compute the range count values corresponding to the point *i* for all the sets *S* containing the dimension *d* in them (there are $2^{d-1}$ such sets), using only the points from the (d-1)-dimensional range tree *RT*. When computing the value $rcount(i,S)$, we will consider only the query ranges for the dimensions $1,\ldots,d-1$ (because all the points in the range $[x(i,d)-D_{cand}, x(i,d)+D_{cand}]$ are already in *RT*. Since the range tree needs to be dynamic (i.e. support point insertions and deletions), we cannot use the fractional cascading technique anymore. Thus, a range count query in this range tree will take $O(log^{d-1}(n))$ time.

If $d=2$ then *RT* may, in fact, be just a segment tree [3]. If $d=1$ then *RT* is just a counter: *RT* is initially *0*, it is incremented by *1* when a point enters the slab, it is decremented by *1* when a point leaves the slab, and its (current) value is the answer for each $rcount(i,\{1\})$ query. The memory consumed by *RT* is $O(n \cdot log^{d-2}(n))$ (or $O(1)$ if $d=1$) and the overall time complexity of the sweeping algorithm is $O(n \cdot log^{d-1}(n)+n \cdot log(n))$. Then, if $d>1$ we will call the algorithm *Sol(d-1)* in order to compute the values $rcount(*,S)$ for the sets *S* not containing the dimension *d*. Thus, the overall time complexity of the algorithm *Sol(d)* is $O(n \cdot (log^{d-1}(n)+log^{d-2}(n)+\ldots+log(n)))= O(n \cdot log^{d-1}(n))$.

The overall time complexity of the algorithms described above is computed by multiplying the time complexity of the feasibility test by that of performing the binary search. Thus, we obtain $O(n \cdot log^{max\{1,d-1\}}(n) \cdot log(D_{cand}))$ algorithms.

However, the time complexity of the feasibility tests can be improved. First, we notice that we can sort the points according to their coordinates in every dimension *j* before binary searching $D_{cand}$. Thus, we will store *d* arrays of sorted points. So far, this only improves the time complexity of the feasibility test in the case $d=1$ (from $O(n \cdot log(n))$ to $O(n)$). Next, during the feasibility test we will sweep the points like before, using the same slab as before (with bounded size in the dimension *d*). We will maintain a (d-1)-dimensional data structure *DS* containing all the points which have already left the slab (*DS* is initially empty). When a point *i* leaves the slab, we add the point $(x(i,1), \ldots, x(i,d-1))$ to *DS*. When a point *i* is at the middle of the slab, we query *DS* to check if:

1. *DS* contains any point *j* with $x(j,k)>x(i,k)+D_{cand}$ (for every $1 \leq k \leq d-1$)
2. *DS* contains any point *j* with $x(j,k)<x(i,k)-D_{cand}$ (for every $1 \leq k \leq d-1$)

If any of the conditions (*1* or *2*) is met for some point *i*, then $D_{cand}$ is a feasible distance. If none of the conditions is met for any point *i*, then $D_{cand}$ is not a feasible distance.

*DS* can easily be implemented as a range tree, but there are other data structures which support this type of range queries (in which the query range is unbounded either towards $+\infty$ or towards $-\infty$) more efficiently. In particular, for $d=2$, *DS* only needs to store the maximum ($x_{max}$) and minimum ($x_{min}$) values of the coordinates $x(j,1)$ of the points *j* which left the slab. Then, if $xmax>x(i,1)+D_{cand}$ then the condition *1* is met for the point *i*, and if $xmin<x(i,1)-D_{cand}$ then the condition *2* is met for the point *i*. $x_{max}$ is initially $-\infty$ and $x_{min}$ is initially $+\infty$. Whenever a point *i* leaves the slab we update $x_{min}$ and $x_{max}$ ($x_{min}=min\{x_{min}, x(i,1)\}$, $x_{max}=max\{x_{max}, x(i,1)\}$). Thus, the time complexity of the feasibility test becomes $O(n)$ for $d=2$ (as the points were sorted before starting the binary search).

V. STRING PROCESSING PROBLEMS

*A. Optimal Encoding using Consecutive Repetitions*

We consider a string *S* composed of *N* characters: $S(1), \ldots, S(N)$. We want to encode the string *S* by using the consecutive contiguous repetitions that occur within *S*. The only operation that we can perform is to replace a sequence of consecutive characters from *S*, of the type $u^k$ (i.e. which is composed of the string *u* repeated *k* times) by the sequence $Codif_1(k) \cdot u \cdot Codif_2(k)$, where by $A \cdot B$ we denote the concatenation of *A* and *B*. $Codif_1(k)$ and $Codif_2(k)$ are two strings marking the be-ginning and the end of the encoded sequence and they do not contain characters which exist in S initially. For instance, the sequence *"abcabcabc"* could be replaced by *"3(abc)"*, if $Codif_1(3)="3("$ and $Codif_2(3)=")"$. The operation of replacing a contiguous subsequence of S can be applied repeatedly, including upon some sequences containing previously encoded subsequences; however, a previously encoded subsequence must be fully included in one of the sequences u of the $u^k$ subsequence being replaced. We want to find a minimum length encoding of S.

In order to solve the problem we will use a dynamic programming approach. We will compute a table $Lmin(i,j)$=the minimum length of encoding the contiguous subsequence of S starting at the position i and ending at the position j. We will compute the values $Lmin(i,j)$ in ascending order of the length of the subsequence (i.e. in increasing order of $j-i+1$). We have $Lmin(i,i)=1$. Then, we iterate with a variable L from 2 to N and, for every value of L, we will first compute all the divisors of L and then we will consider every possible value of i ($1 \leq i \leq N-L+1$). Once we considered the value i, the value of j is equal to $i+L-1$. We will initialize $Lmin(i,j)=j-i+1$. Then, we will consider all the divisors k of L and we will verify if the contiguous subsequence $S(i)S(i+1)...S(j)$ has the structure $u^k$. The verification can be performed in linear time: we will consider every position p from $i+L/k$ to j and, if $S(p) \neq S(p-L/k)$ (for at least one position p), then the considered subsequence does not have the structure $u^k$; otherwise, it has this structure. If the subsequence has the desired structure, then we will set $Lmin(i,j)=min\{Lmin(i,j), len(Codif_1(k)) + len(Codif_2(k)) + Lmin(i,i+L/k-1)\}$. The overall time complexity is $O(N^{3.5})$.

### B. Counting the Number of Occurrences of a Substring in a String Generated by a Grammar

We consider a grammar consisting of a set of terminal symbols T and N non-terminal symbols, numbered from 1 to N. Each non-terminal symbol X is specified as a sequence consisting of L(X) symbols: $R(X,1), ..., R(X,L(X))$. Each symbol is either a terminal symbol or a non-terminal symbol Y ($1 \leq Y \leq X-1$). Let's consider the string $Exp(N)$, obtained by expanding the non-terminal N (i.e. a string consisting of terminal symbols only). We are given a string S and we want to compute the number of occurrences of S in $Exp(N)$. Two occurrences of S may partially overlap.

Computing the string $Exp(N)$ and then searching for the string S in $Exp(N)$ (in $O(|Exp(N)|)$ time) is not possible, because the string $Exp(N)$ may be exponentially large. Instead, we need an approach which does not compute $Exp(N)$.

We will start by computing the (KMP) prefix function for the string S. Let $S(1), ..., S(LS)$ be the characters of the string S (LS=the length of the string S). Let P(i) be the prefix function value for the prefix 1, ..., i. We can compute the prefix function in $O(LS)$ time. Then, we define by $Pexp(i,k)$=the string obtained by expanding the sequence composed of the symbols $R(i,k), ..., R(i,L(i))$ ($1 \leq i \leq N$; $1 \leq k \leq L(i)+1$). We will compute the values $C(i,j,k)$=the number of occurrences of S in $Pexp(i,k)$, if we consider that the first j characters of S are already "matched" before starting $Pexp(i,k)$. Thus, when $Pexp(i,k)$ begins, the prefix of length j of the string S is already matched over the j characters which are located right before $Pexp(i,k)$ in the final string. We will also compute $Suf(i,j,k)$=the longest (incomplete) prefix of S which matches the end of the string $Pexp(i,k)$, under the same conditions as above (i.e. the first j characters of S match the j characters located right before the start of $Pexp(i,k)$).

We will consider the non-terminal symbols in increasing order: $i=1,...,N$. We have $C(i,j,L(i)+1)=0$ and $Suf(i,j,L(i)+1)=j$ ($0 \leq j \leq LS-1$). We will then consider the positions k, in decreasing order, starting from L(i) and ending with the position 1 ($k=L(i), L(i)-1, ..., 1$). If $R(i,k)=q$, where q is a non-terminal symbol, then we will proceed as follows: *(1)* we initialize $ne=Suf(q,j,1)$; *(2)* we set $C(i,j,k)=C(q,j,1)+C(i,ne,k+1)$; *(3)* $Suf(i,j,k)=Suf(i,ne,k+1)$. If $R(i,k)$ is a terminal symbol, then let $ne=j$. While ($ne>0$) and ($S(ne+1) \neq R(i,k)$) we set $ne=P(ne)$. At the exit of the *while* loop, if $S(ne+1)=R(i,k)$, then we increment ne by 1. If $ne=LS$, then we set $caux=1$ and $ne=P(ne)$; otherwise, we set $caux=0$. Then we will have $C(i,j,k)=caux+C(i,ne,k+1)$ and $Suf(i,j,k)=Suf(i,ne,k+1)$. The final answer is $C(N,0,1)$.

## VI. CONCLUSIONS AND FUTURE WORK

In this paper we presented novel algorithmic techniques for several multidimensional (and 1D) data processing problems, related to clustering, encoding and the computation of the diameter using a new distance function. All the presented algorithms were thoroughly analyzed from a theoretical point of view, and some of them even from a practical point of view.

All the presented problems are interesting from the point of view of multidimensional data analysis, while some of them have applications in the field of distributed computing (e.g. the problem presented in Section III).

As future work, we intend to implement and evaluate all the proposed solutions (so far, only some of them have full implementations) and continue the study of multidimensional data processing problems in need of novel efficient algorithmic techniques.